# Nanoscale confinement of all-optical switching in TbFeCo using plasmonic antennas

TianMin Liu, Tianhan Wang, Alexander H. Reid*, Matteo Savoini, Xiaofei Wu, Benny Koene, Patrick Granitzka, Catherine Graves, Daniel Higley, Zhao Chen, Gary Razinskas, Markus Hantschmann, Andreas Scherz, Joachim Stöhr, Arata Tsukamoto, Bert Hecht, Alexey V. Kimel, Andrei Kirilyuk, Theo Rasing & Hermann A. Dürr.

**All-optical switching (AOS) of magnetic domains by femtosecond laser pulses was first observed in the transition metal-rare earth (TM-RE) alloy GdFeCo[1-5]; this phenomenon demonstrated the potential for optical control of magnetism for the development of ever faster future magnetic recording technologies. The technological potential of AOS has recently increased due to the discovery of the same effect in other materials, including RE-free magnetic multilayers[6,7]. However, to be technologically meaningful, AOS must compete with the bit densities of conventional storage devices, restricting optically-switched magnetic areas to sizes well below the diffraction limit. Here, we demonstrate reproducible and robust all-optical switching of magnetic domains of 53 nm size in a ferrimagnetic TbFeCo alloy using gold plasmonic antenna structures. The confined nanoscale magnetic reversal is imaged around and beneath plasmonic antennas using x-ray resonant holographic imaging. Our results demonstrate the potential of future AOS-based magnetic recording technologies.**

The first demonstration of all-optical switching (AOS) in the rare-earth–transition metal alloy $Gd_xFe_yCo_{100-x-y}$ by Stanciu et al.[1] showed switched areas with a diameter of 10 μm, defined by the laser spot size used. Because of its magnetic softness, the minimum stable domain size in a continuous GdFeCo film is about 1 μm. All-optical switching was further localized in GdFeCo down to 200 nm thanks to sample patterning[8,9], while in the magnetically harder TbFeCo, typical bit sizes down to 300 nm are achieved by focusing the light with a microscope objective[10] and exploiting the AOS-threshold character[11]. However, further shrinking of the switching domain sizes towards the $Tbit/inch^2$ densities of projected magnetic storage devices requires the use of a different approach.

In order to achieve unprecedented nanoscale control of AOS switching areas, we employ two-wire plasmonic gold nano-antennas to obtain the required localized enhancement of the optical field around the structure[12,13]. The two-wire antenna used consists of two arms with a narrow gap. By placing this plasmonic resonator in close proximity of the active magnetic layer, we exploit the near-field enhancement to confine the switched area to dimensions dictated by the extension of the antenna near-field. In order to resolve the resulting magnetic domain, we use an x-ray holographic imaging technique, which has a resolving power down to 16 nm[14]. The magnetic contrast is obtained by using resonant circularly polarized x-rays through x-ray magnetic circular dichroism (XMCD).

The sample stack, shown in Fig. 1a. A 20 nm TbFeCo magnetic layer is sputtered onto a 100 nm thick $Si_3N_4$ membrane, and is sandwiched between 10 nm (top side) and 5 nm (bottom side) of in-situ grown $Si_3N_4$ to prevent oxidation (see methods). The bottom of the membrane is covered by 800 nm of gold, and a milled-out imaging window of 2x2 µm² is produced using focused ion beam (FIB). An L-shaped x-ray reference slit is also milled in proximity of the imaging window, but extends throughout all sample layers.

The optimum antenna geometry is determined by performing finite difference time domain simulations (FDTD), considering the sample stack, laser wavelength, polarization and other experimental parameters (see methods). Three antennas sets were prepared with total lengths of 230, 270 and 310 nm and antenna gaps of between 20 to 30 nm (see Fig. 1b). The single-crystalline gold antennas were first fabricated by FIB milling on a glass substrate coated with gold and $SiO_2$ layers before being transferred to the current sample (see methods). This fabrication procedure prevents damage to the magnetic material underneath the antennas by FIB milling[16]. Magnetic switching is achieved by directing single fs optical pulses onto the sample surface that contains the antennas. The optical beam focus is 75 µm full-width-half-maximum (FWHM), and we can therefore consider the antennas within the imaging window to be excited by a plane wave. After the optical exposure, the magnetic state is probed using a resonant circularly polarized x-ray beam from the Stanford Synchrotron Radiation Lightsource (SSRL) in transmission geometry. The diffraction pattern is formed by interference between the transmitted radiation through the imaging window and the reference slit, and is recorded on a CCD detector (see Fig. 2). To reconstruct the x-ray transmission image, we use holography with extended reference by autocorrelation linear differential operation (HERALDO)[14,15] (see methods).

The x-ray diffraction experiment is performed at the SSRL beamline 13-3, with the incident x-rays tuned to Fe $L_3$ edge at 706.8 eV. The x-ray transmission profile obtained from the HERALDO reconstruction resolves both magnetic and non-magnetic absorption contrasts (see Fig. 3a & 3b). To isolate the magnetic information, we subtract the reconstructed images obtained with right- and left-circularly polarized x-rays. This difference reveals the magnetic domain structure of the sample, as shown in Fig. 3c. The sources of non-magnetic contrast, such as the gold antennas observed in Fig. 3a & 3b, disappear from the difference image Fig. 3c. The non-magnetic contrast determines the position of the magnetically switched areas with respect to the antennas.

Figure 4 shows our demonstration of AOS at the nanoscale. A reversed magnetic domain is written in a uniformly magnetized region below a 230 nm antenna arm by a single optical pulse of 3.7 mJ/cm² (Fig. 4a, 4b). The diameter of the switched area is 53 nm, almost six times smaller than previous observations[10]. The observed magnetic switching by this plasmonic enhancement is both reversible and reproducible, as shown in subsequent panels of Fig. 4. Areas with switched magnetic orientation are observed below the antenna. The switched domain size of 53 nm corresponds to an optical localization of about λ/20. This domain size compares favorably to the tracks widths of 55 nm used in a recent demonstration of 1+ Tb/in² heat assisted magnetic recording[17].

The magnetic state can be toggled back and forth deterministically, demonstrating that the optically-induced switching is controllable (see Fig. 4). Following the writing of the initial switched domain in Fig. 4b, the sample is then magnetically reset ex-situ by applying a magnetic field to the sample (Fig. 4c), returning the magnetization to the initial state. Subsequently, it receives another laser pulse at a slightly higher fluence, of 4 mJ/cm²—this fluence change is due to the necessary x-ray – optical realignment subsequent to the magnetic reset. Magnetic reversal is again observed in the same region (Fig. 4d), demonstrating that the switching occurs reproducibly. However, the area and shape of the switched region is not identical. Illuminating the sample again with a further laser pulse, identical to the latter one of 4 mJ/cm$^2$, the magnetic state is returned back to its original uniformly magnetized state (see Fig. 4e).

The optimal conditions for nanoscale all-optical switching are found by varying the optical laser fluence and observing the induced magnetic domain structures, as shown in Fig. 5. For each fluence a new, pristine sample is used. The HERALDO images show that with a fluence of 3.7 mJ/cm$^2$, magnetic domains of reversed sign (red in Fig. 5) are written into the original uniform (blue) magnetization in the vicinity of two antennas. Below this threshold no magnetic switching takes place; while at higher fluences, above 10.3 mJ/cm$^2$, switching also occurs in areas where no near-field enhancement is expected. From these observations we estimate the effective near-field enhancement of AOS by the antennas in the TbFeCo magnetic layer to be about 2.5. This enhancement is comparable to the simulated intensity enhancement for the 230 nm antenna of between 3 and 5 shown in Fig 1b. . At much higher fluences, a magnetic multi-domain state is observed, for example, see in Fig. 3c which used 21 mJ/cm$^2$. At this fluence condition any AOS is overcome by thermal fluctuations after the sample is heated above the magnetic ordering temperature.

Although the switching is reproducible, we do not yet have full control over the magnetic switching location. Comparing the simulated field profile depicted in Fig. 1b to the experimental observed switching locations highlights the question of location control. Shaping of the excitation laser pulse offers a path to better control of the near-field AOS process[18]. Although the FDTD simulation indicates that there is a relatively homogenous area of laser intensity enhancement, the magnetic switching only occurs at the end of one antenna arm in Fig 4. Fabrication imperfections, surface roughness and/or a non-perfect positioning of the antennas might break the illumination symmetry, but the SEM measurements reveal little evidence of such imperfections.

Another possible reason for the distribution in switching-event locations is an inhomogeneous distribution of the switching fluence threshold in the TbFeCo layers due to inhomogeneity in the Tb, Fe and Co composition in the alloy. Such a difference has been reported for GdFeCo[19]. In the TbFeCo samples investigated here, the switching of small isolated areas away from antennas and aperture edges after illumination with pulses with fluencies above 9 mJ/cm$^2$, supports the suggestion of a spatially-inhomogeneous switching threshold. Spatially resolved energy dispersive x-ray spectroscopy of a TbFeCo film that was co-sputtered with the measured sample shows nanoscale chemical inhomogeneities. We find variations in the local Tb and Fe concentrations by up to 7 and 8%, respectively over ~10 nm length scales (see supplementary materials). Switching in GdFeCo has been theoretically shown to be strongly concentration dependent[20] and a similar behavior is expected for TbFeCo.

At this stage, it is very difficult to further comment on the switching yield differences between the different antenna sizes, in view of the measured local variations in magnetic properties and switchability. In order to investigate the reliability of the simulated field intensities and to provide a thorough study of maximum achievable localization of the all-optical switching, we would need a material showing a more uniform composition (thus switching threshold) such as multilayered samples[6].

In conclusion, by exploiting the field-confining properties of plasmonic nanoantennas, we successfully demonstrated all-optical switching with a lateral size of 53 nm, in a ferrimagnetic TbFeCo thin film, a size comparable with that achieved in heat assisted magnetic recording. To visualize the switching we used resonant x-ray holography. The switching is shown to be reproducible and back and forth switching was demonstrated. This achievement indicates the compatibility of AOS with optical near-field techniques and its potential as a magnetic recording technology. The method of study is further compatible with a dynamical study of the AOS process[21]. Our results, moreover, show the necessity of a better engineering of the sample structure; a more homogenous distribution of the elements is a key step for AOS-based technological applications. This might be more easily achieved in multilayer systems.

**Acknowledgements:**

Research at Stanford is supported by US DOE, Office of Basic Energy Sciences, Materials Sciences and Engineering Division under contract DE-AC02-76SF00515. Portions of this research were carried out at the Stanford Synchrotron Radiation Lightsource, a Directorate of SLAC National Accelerator Laboratory and an Office of Science User Facility operated for the US Department of Energy Office of Science by Stanford University. Furthermore this research has received funding from Stichting voor Fundamenteel Onderzoek der Materie (FOM), De Nederlandse Organisatie voor Wetenschappelijk Onderzoek(NWO), the European Union (EU) Nano Sci-European Research Associates (ERA) project FENOMENA, ERCGrant agreement No. 257280 (Femtomagnetism) and No.339813 (EXCHANGE) and EC FP7 No. 281043 (FEMTOSPIN).

**Affiliations:**

*Stanford Institute for Materials and Energy Sciences, SLAC National Accelerator Laboratory, 2575 Sand Hill Road, Menlo Park, CA 94025, USA.*

TianMin Liu, Tianhan Wang, Alexander H. Reid, Patrick Granitzka, Catherine Graves, Daniel Higley, Zhao Chen, Andreas Scherz, Joachim Stöhr & Hermann A. Dürr.

*Institute for Molecules and Materials, Radboud University Nijmegen, Heyendaalseweg 135, 6525 AJ Nijmegen, The Netherlands.*

Matteo Savoini, Benny Koene, Alexey V. Kimel, Andrei Kirilyuk & Theo Rasing

*Nano-Optics and Biophotonics Group, Experimentelle Physik 5, Physikalisches Institut, Wilhelm-Conrad-Röntgen-Center for Complex Material Systems, Universität Würzburg, Am Hubland, Würzburg D-97074, Germany.*

Xiaofei Wu, Gary Razinskas & Bert Hecht.

*Institute Methods and Instrumentation for Synchrotron Radiation Research, G-ISRR, Helmholtz-Zentrum Berlin, Albert-Einstein-Str 15, 12489 Berlin, Germany.*

Markus Hantschmann.

*College of Science and Technology, Nihon University, 7-24-1 Funabashi, Chiba 274-8501, Japan*

Arata Tsukamoto.

**Author Contributions:**

A.H.R. & M.S. conceived the experiment. B.K., M.S., X.W., T.W. & B.H. designed the antennas. A.T. grew the TbFeCo films. X.W. milled and transferred the antennas. T.W. & T.L. milled the references. T.W., A.H.R., T.L., P.G. & M.S. constructed the experiment. T.W., A.R., M.S., B.K., P.G., C.G., D.H., Z.C., & M.H. performed the measurements and online analysis. T.L., T.W., H.D. & A.R. performed the offline analysis. B.K., G.R. & M.S. performed the FDTD simulations. A.R., T.L., M.S., B.K., A.K., T.R. & H.A.D. co-wrote the manuscript with input from all authors.

**Methods:**

**Samples:** 20 nm films of $Tb_{22}Fe_{69}Co_9$ are fabricated by magnetron sputtering onto 100 nm $Si_3N_4$ membranes. The sample is sandwiched between 2 extra protective layers of $Si_3N_4$ respectively of 5 nm (bottom) and 10 nm (top). The samples exhibit out-of-plane magnetization with a coercive field of 0.65 T at room temperature. The Curie temperature is at 550K, while the magnetization compensation temperature, i.e. the temperature at which the ferrimagnetic alloy behaves as a pure antiferromagnet, is below room temperature.

To fabricate the nanoantennas, first single crystalline gold flakes grown on a glass coverslip were transferred to another glass coverslip coated with 50 nm gold and 40 nm $SiO_2$ layers sequentially[22]. Then nanoantenna arrays were milled out of the gold flakes with focused ion beam (FIB). Afterwards a Poly(methyl methacrylate) (PMMA) layer (about 300 nm) was spin-coated on the sample and then baked in an oven at 170 °C for 2 hours. Due to the poor adhesion between the gold layer and coverslip and the hydrophobicity of PMMA, the gold/SiO2/PMMA layers and the sandwiched antennas were peeled off from the coverslip by dipping the sample into water obliquely because of the surface tension of water. After etching away the gold and $SiO_2$ layers with $KI/I_2$ aqueous solution and buffered oxide etch respectively, the antennas embedded in the PMMA layer were placed on top of the membranes with help of micro-manipulators under an optical microscope. As the last step, the PMMA layer was dissolved in acetone vapour.

The 800-nm-thick gold layer is sputtered on the backside of the sample. Imaging aperture and reference slots are fabricated with a FIB of 20 nm focus. The distance between this slit and the imaging window is within the x-ray spatial coherence length in order to fulfill the necessary conditions for holographic imaging[14].

**Experiment:** The X-ray scattering experiment is performed at Stanford Synchrotron Radiation Lightsource (SSRL) beamline 13-3. A 1030 nm Calmar Cazadero Er-doped fiber laser is operated in pulse mode at 320 kHz with 500 fs long pulses of 8 µJ/pulse. Single pulses are selected from the train by use of an acousto-optic modulator and manual shutter. The laser pulse is S-polarized with the electric field vector along the long axis of the dipole antennas.

The incident circularly polarized X-ray has a spot size of 220x70 µm² and is tuned to the Fe $L_3$ edge to probe the Fe atoms selectively. To improve the x-ray beam coherence a 100 µm aperture is placed 600 mm upstream of the sample position.

The x-ray diffraction pattern is then collected using an in-vacuum CCD camera at a distance of 200 mm away from the sample plane. The camera comprised of 2048 by 2048 pixels of 13.5 µm in size. A beamstop of approximately 0.5 mm in radius is used to block the undiffracted beam.

Holographic reconstructions are performed by the method of holography with extended reference by autocorrelation linear differential operation (HERALDO). The reference is an L-shaped slit milled approximately 5 µm away from the imaging aperture. To reconstruct the absorption contrast x-ray image, a linear operator, whose inverse Fourier transform (IFT) is a differential operator, is applied to the diffraction image.

**Simulations:** The simulations are Finite Difference Time Domain (FDTD) simulations performed with the commercial software Lumerical FDTD[23]. The simulation consists of an area of $1 \times 1 \times 1.1 \mu m^3$ with a non-uniform meshing with the smallest mesh cell being $1\ nm^3$ around the antenna structure. The boundary conditions consist of perfectly matched layers and we make use of the symmetry of the structure to reduce the calculation time. The dielectric constants used for the different materials are $\varepsilon = -45.1 + 3.25i$ (ref 24) for gold, $\varepsilon = 4$ (ref 25) for $Si_3N_4$ and measured to be $\varepsilon = 23.6 + 41.3i$ for TbFeCo. The antenna is modelled to match the fabricated structure as realistically as possible by rounding its corners. The height and the width at the base of the antenna are both 55 nm. The gap at the base is 20 nm. A plane wave source at a wavelength of 1030 nm is used for excitation.

**Figures**

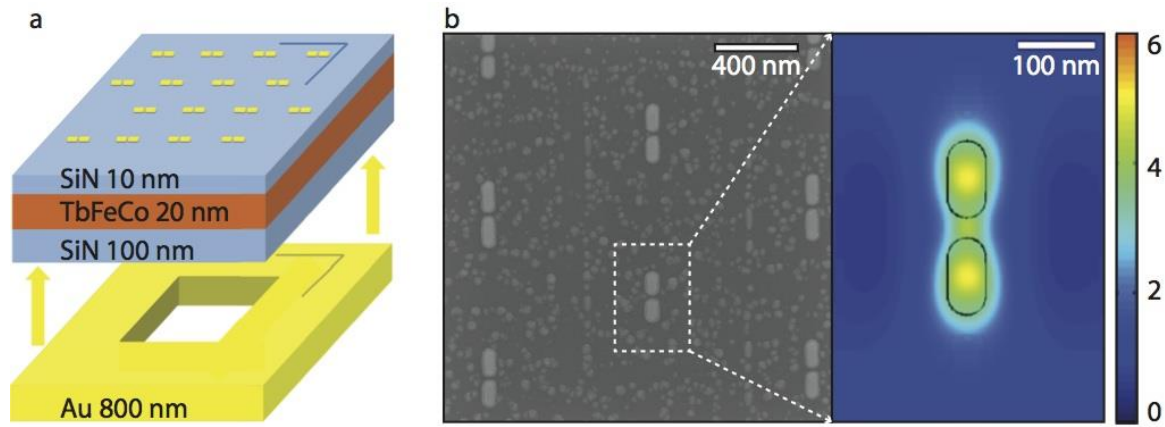

**Figure 1 | Plasmonic antennas combined with magnetic TbFeCo layer for nanoscale optical enhancement. a,** schematics of the sample structure showing the depth profile (not to scale). Gold nano-antennas are positioned onto the topmost surface, a 10 nm $Si_3N_4$ capping layer which protects the magnetic TbFeCo layer from oxidation. The reverse side is covered by an 800 nm gold layer, in which a $2 \times 2\ \mu m^2$ imaging window is milled. An L-shaped holographic reference slot is cut through all layers. **b,** a scanning electron microscope image of the gold antenna structures on the top surface of the sample. Three different antenna lengths of 230, 270, and 310 nm are present. The simulated near-field intensity enhancement profile for the 230 nm antenna is shown on the right. The near-field enhancement is simulated at the middle of the TbFeCo layer using a plane wave with a wavelength of 1030 nm for excitation. The color scale shows that underneath the antenna the field intensity is enhanced up to approximately five times the incident intensity.

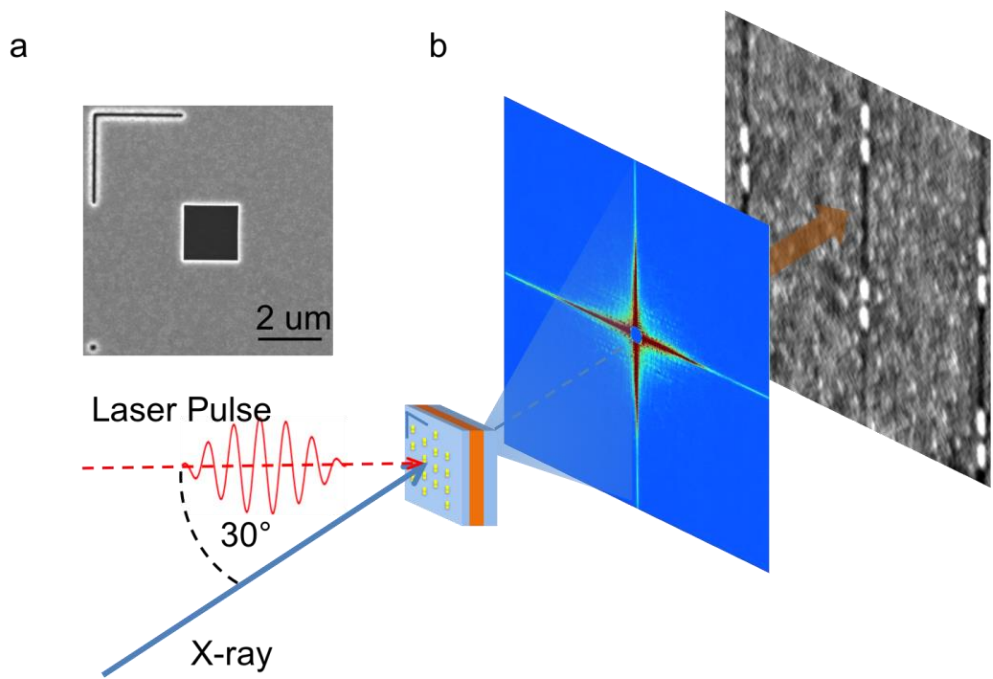

**Figure 2 | Optical switching and resonant x-ray holography schematic**. **a,** an SEM image of the gold holographic mask on the backside of the sample, with the center imaging window, L-shaped reference slit and a point reference. **b,** Experiment schematic: a 500 fs incident laser pulse (center wavelength 1030 nm and linearly polarized along the antennas long axes) induces the optical switching. This pulse is incident onto the sample at 30 degrees angle from normal. X-ray diffraction patterns are collected using a normally incident beam of right- and left-circularly polarized x-rays from the SSRL tuned to the Fe $L_3$ resonance at 706.8 eV. The reconstructed image of the sample is obtained from the inverse Fourier transform of the diffraction image after filtering[14,15].

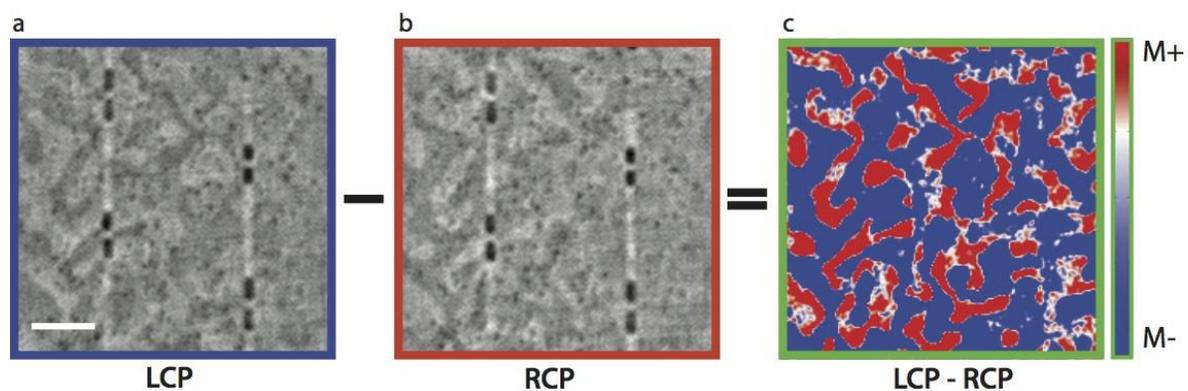

**Figure 3 | Separating magnetic and non-magnetic image contrast. a, b**, Individual holographic image reconstructions obtained with LCP (left) and RCP (middle) x-rays at 706.8 eV, show a superposition of magnetic domains and the antenna structures. The white scale bar in figure **a** is 400 nm in length. **c**, The difference image (right), LCP – RCP, reveals the magnetic domain structure. A multi-domain magnetic structure was prepared by using an incident laser fluence of 21 mJ/cm$^2$.

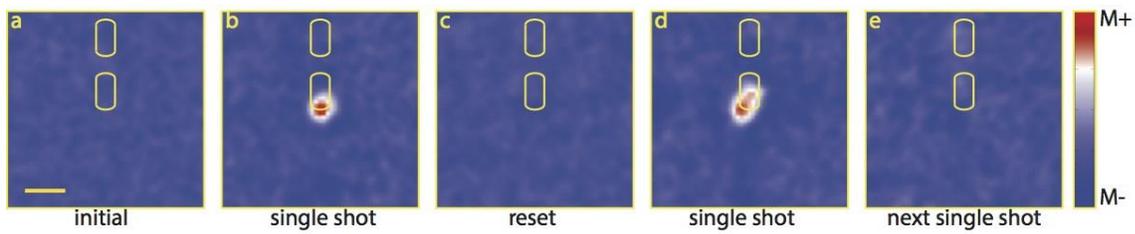

**Figure 4 | Antenna-mediated switching is reproducible and reversible. a,** initial magnetic contrast around the selected antenna after being magnetic saturated in a field of 1.6 T. The gold scale bar is 100 nm in length. **b,** magnetic contrast after 1$^{st}$ laser pulse of 3.7 mJ/cm$^2$. A small domain with a FWHM of 53.4 nm is switched. **c,** the magnetization is reset again using an external magnetic field. **d,** magnetic contrast after the 1$^{st}$ laser pulse of 4.0 mJ/cm$^2$ on the newly saturated sample. A domain of comparable size as in **b** is switched in the same region. **e.** Magnetic contrast after a 2$^{nd}$ laser pulse of 4.0 mJ/cm$^2$. The magnetization of the region switched in **d** is toggled back to its original state.

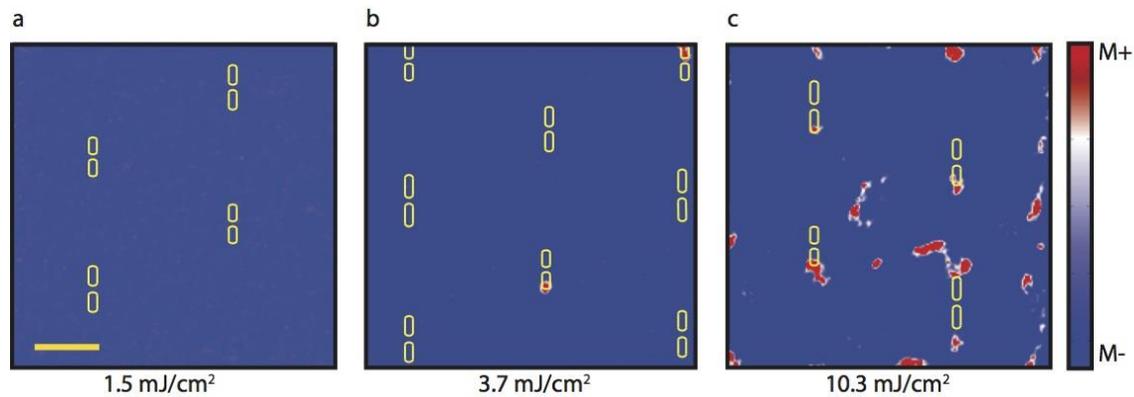

**Figure 5 | Isolated magnetic switching at different laser fluences.** Each panel shows the reconstructed magnetic contrast image with the antenna positions outlined. The gold scale bar in figure **a** is 400 nm in length. **a,** Below 3.7 mJ/cm² incident fluence, laser pulses induce no magnetic changes, and the magnetization remains in its single-domain initial state. **b,** Isolated magnetic switching occurs underneath two antenna structures at 3.7 mJ/cm² incident fluence, yielding the fluence threshold for antenna enhanced switching. **c,** At higher fluences of 10.3 mJ/cm², switching is observed in multiple locations, including around all antennas and near the edges of the imaging aperture. The aperture-edge switching is caused by intensity enhancements from back reflections of the optical pulses from the imaging window edge. FDTD simulations show that this reflection results in an intensity enhancement of about 1.7. Uniform switching of the entire imaged area is not observed at any fluence. At fluences above roughly 20 mJ/cm², the induced final state is a random multi-domain state not correlated to the initial state. From these measurements we can estimate the effective intensity enhancement of the antennas to be about 2.5.

**References**

1. Stanciu, C. D. *et al.,* All-optical magnetic recording with circularly polarized light. *Phys. Rev. Lett.* **99,** 047601 (2007).
2. Radu, I. *et al.* Transient ferromagnetic-like state mediating ultrafast reversal of antiferromagnetically coupled spins. *Nature* **472,** 205 (2011).
3. Ostler, T. A. *et al.* Ultrafast heating as a sufficient stimulus for magnetization reversal in a ferrimagnet. *Nature Commun.* **3,** 666 (2012).
4. Vahaplar, K. *et al.* Ultrafast Path for Optical Magnetization Reversal via a Strongly Nonequlibrium State. *Phys. Rev. Lett.* **103,** 117201 (2009).
5. Steil, D., *et al.,* All-optical magnetization recording by tailoring optical excitation parameters. *Phys. Rev. B,* **84**, 224408 (2011).
6. Mangin, S. *et al.* Engineered materials for all-optical helicity-dependent magnetic switching. *Nature Mater.* **13,** 286 (2014).
7. Lambert, C-H. *et al.* All-optical control of ferromagnetic thin films and nanostructures. *Science* DOI:10.1126/science.1253493 (2014).
8. Le Guyader, L. *et al.* Demonstration of laser induced magnetization reversal in GdFeCo nanostructures. *Appl. Phys. Lett.* **101,** 022410 (2012).
9. Savoini, M. *et al.* Highly efficient all-optical switching of magnetization in GdFeCo microstructures by interference-enhanced absorption of light. *Phys. Rev. B,* **86** 140404(R) (2012).
10. Finazzi, M. *et al.*, Laser-induced magnetic nanostructures with tunable topological properties, *Phys. Rev. Lett*., **110,** 177205 (2013).
11. A. R. Khorsand *et al.,* Role of magnetic circular dichroism in all-optical magnetic recording, *Phys. Rev. Lett.,* **108**, 127205 (2012).
12. P. Mühlschlegel *et al.,* Resonant optical antennas, *Science* **308,** 1607 (2005).
13. B. Koene *et al.* Optical energy optimization at the nanoscale by near-field interference, *App. Phys. Lett.* **101,** 013115 (2012)
14. Zhu, D. *et al.* High-Resolution X-Ray Lensless Imaging by Differential Holographic Encoding. *Phys. Rev. Lett.* **105**, 043901 (2010).
15. Guizar-Sicairos, M. and Fienup, J. R., Holography with extended reference by autocorrelation linear differential operation. *Opts. Express,* **15,** 17592 (2007).
16. M. Savoini *et al.* Attempting Nanolocalization of All-Optical Switching through Nano-holes in an Al-mask. *Proc. Of SPIE,* 9167-84 (2014).
17. Wu, A. Q. *et al.,* HAMR Areal Density Demonstration of 1+ Tbpsi on Spinstand. *IEEE Trans. Magn.,* **49,** 779 (2013).
18. Aeschlimann, M., *et al.,* Adaptive subwavelength control of nano-optical fields. *Nature* **466,** 301 (2007).
19. Graves, C. E. *et al.,* Nanoscale spin reversal by non-local angular momentum transfer following ultrafast laser excitation in ferrimagnetic GdFeCo. *Nature Mater.* **12,** 293 (2013).
20. Barker, J., *et al*., Two-magnon bound state causes ultrafast thermally induced magnetization switching, *Scientific Reports* **3**, 3262, DOI: 10.1038/srep03262 (2013).
21. Von Korff Schmising, *et al.,* Imaging Ultrafast Demagnetization Dynamics after a Spatially Localized Optical Excitation, *Phys. Rev. Lett.,* **112,** 217203 (2014).
22. Jiao *et al.,* Creation of nanostructures with poly-mediated nanotransfer printing, *JACS*, **130,** 12612, (2008).

23. FDTD Solutions v8.5.3, 2012, Lumerical Solutions Inc., Vancouver, Canada.
24. Johnson, P. B. and Christy, R. W. Optical constants of the noble metals, *Phys. Rev. B*, **6,** 4370 (1972).
25. Palik, E. D. Handbook of Optical Constants of Solids I-III (Academic Press, 1997).

# Supplementary Materials

1. **Plasmonic field enhancement depending on antenna size**

To find the right design for the antennas that had to be fabricated we first performed Finite Difference Time Domain simulations (FDTD). The basic simulation method is described in the methods section of the main article. From the simulations we determined the optimal antenna length to be 270 nm. This antenna length gave the largest intensity enhancement (about 8 times) in the center of the TbFeCo layer directly below the antenna gap. Here the intensity enhancement is defined as the intensity profile obtained with the antenna structure divided by the intensity profile obtained without the antenna structure. To take care of possible differences between the real optimal antenna length and the simulated one, we fabricated antennas with a total length of 230 270 and 310 nm. The intensity profiles of these antennas at half height of the TbFeCo layer are shown in Figure S1, and the intensity-depth profiles are shown in Figure S2.

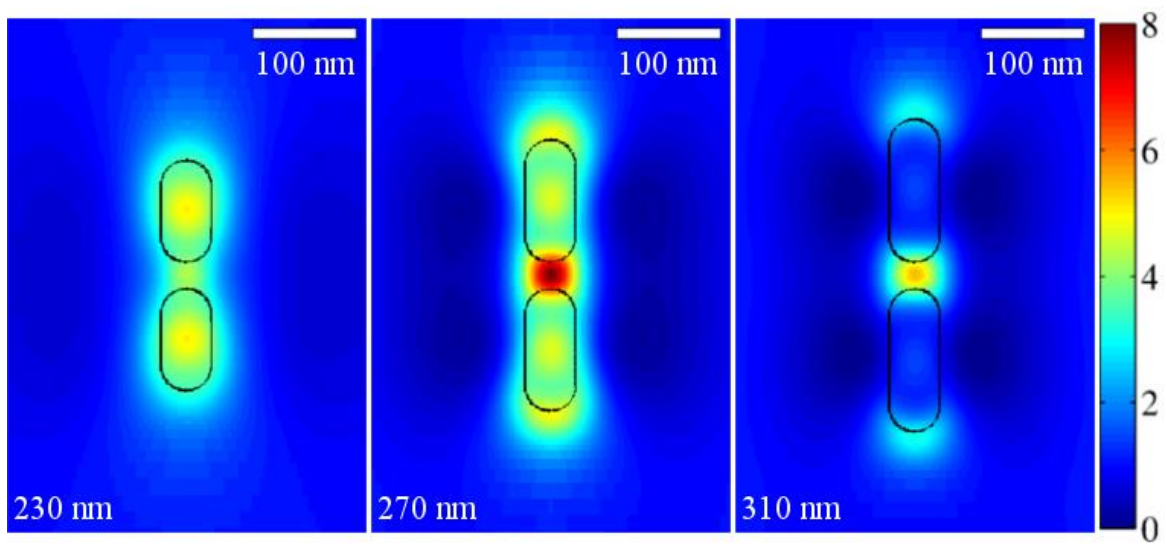

**Figure.S1: FDTD simulation of enhancement effects of two-wire antennas with three different sizes: 230, 270 and 310 nm.** The enhancement is shown as the field intensity at the middle of TbFeCo layer normalized by the case without antennas. White bar at top right corner is 100 nm in length.

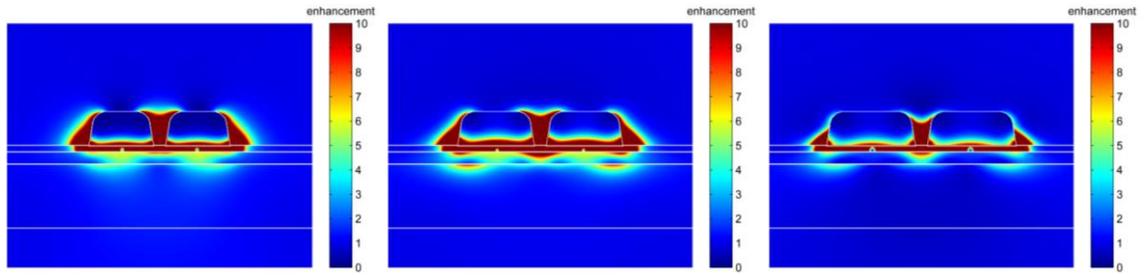

**Figure S2 | FDTD simulation normal to the sample plane of enhancement effects of the two-wire antennas with three different sizes: 230, 270 and 310 nm.** The enhancement is shown for the field intensity at the middle of the respective antennas normalized by the case without antennas.

Although the simulations showed the largest intensity enhancement in the gap of the 270 nm antenna, in the experiments we observed the initiation of AOS at low fluences occurs most often underneath one of the arms of the 230 nm antenna and not in the center. None of the profiles in Figures S1, S2 can explain this asymmetric switching.

2. Sample TEM characterization

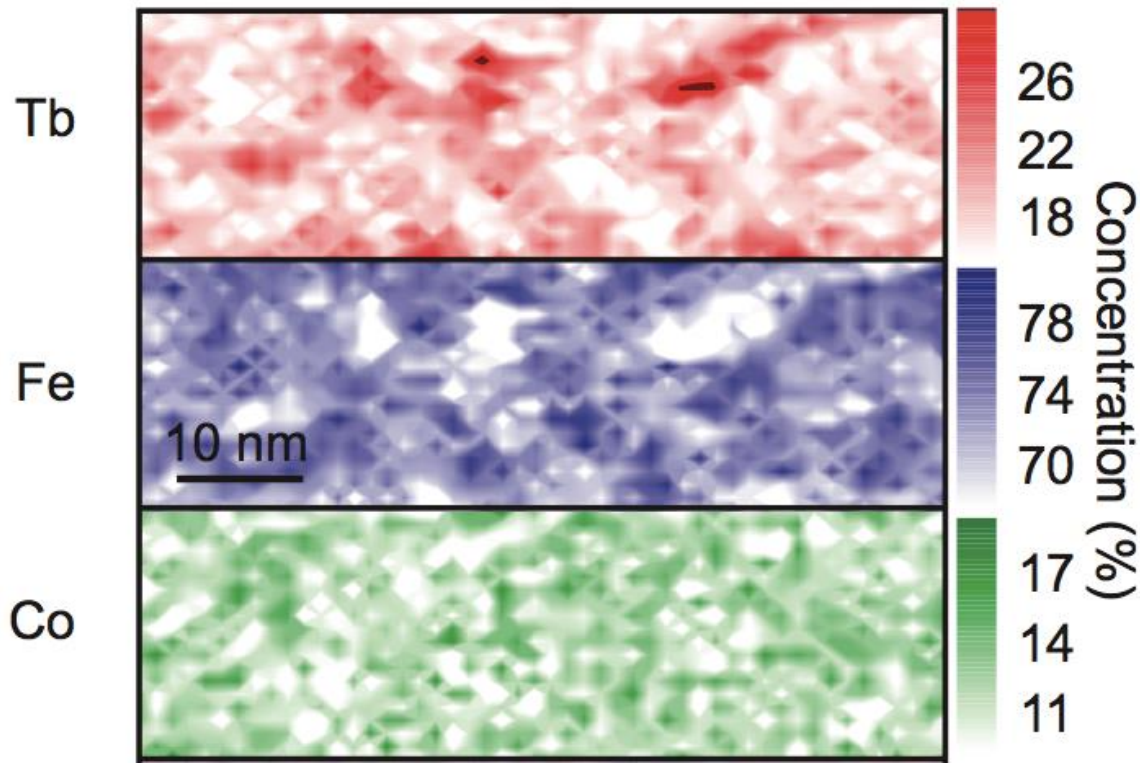

**Figure S3 | STEM-EDAX elemental mapping of the TbFeCo sample.** Darker and lighter regions represent higher and lower concentrations. Nanoscale inhomogeneity is present as shown by the Fe-rich and Tb-rich regions which anticorrelate with each other.

The nanoscale inhomogeneity is imaged in real space with an energy dispersive spectrometry (EDXEDX) detector in scanning transmission electron microscopy (STEM) mode, as shown in Figure S3. There are clear Tb-rich and Fe-rich regions which are anti-correlated with each other. The measured average concentration of $Tb_{16.9}Fe_{71.8}Co_{11.3}$ agrees reasonably well with the expected values. The STEM-EDX and AOS switching TbFeCo samples are deposited in the same run but on different substrates. The TEM $Si_3N_4$ substrate is thinner at 20 nm compared to 100 nm for those used in the switching experiment. A FEI Tecnai G2 F20 X-TWIN TEM with an EDAX SUTW (super ultrathin window) and analyzer is used for elemental mapping.

From resonant magnetic scattering measurements done on GdFeCo [ref main paper], one can reasonably conclude that the charge and magnetic orderings in TbFeCo are similarly correlated. It is expected that the variations in the elemental composition of the alloy will directly influence the local compensation and Curie temperatures of the sample, and vary its magnetic properties. Correspondingly, the local AOS switching threshold will also be affected.